\documentclass[twocolumn]{aastex63}
\usepackage{amsmath,amstext,amssymb}
\usepackage{xspace}
\usepackage{lipsum}
\newcommand{\Milwaukee}{University of Wisconsin-Milwaukee, Milwaukee, WI 53201, USA}
\newcommand{\UTokyo}{Institute for Cosmic Ray Research, The University of Tokyo, 5-1-5 Kashiwanoha, Kashiwa, Chiba 277-8582, Japan}
\newcommand{\Cardiff}{Cardiff University, Cardiff CF24 3AA, UK}
\newcommand{\CMU}{McWilliams Center for Cosmology, Department of Physics, Carnegie Mellon University, Pittsburgh, PA 15213, USA}

\newcommand{\GW}{GW190425\xspace}
\newcommand{\FRB}{FRB20190425A\xspace}
\newcommand{\GAL}{UGC10667\xspace}
\newcommand{\low}{low spin\xspace}
\newcommand{\high}{high spin\xspace}

\shortauthors{Bhardwaj et al.}

\begin{document}

\title{Challenges for Fast Radio Bursts as Multi-Messenger Sources from Binary Neutron Star Mergers} %\GW and \FRB: 
\shorttitle{\GW and \FRB}

\author[0000-0002-3615-3514]{Mohit Bhardwaj}
\affiliation{\CMU}
\email{mohitb@andrew.cmu.edu}

\author[0000-0002-6011-0530]{Antonella Palmese}
\affiliation{\CMU}

\author[0000-0003-2362-0459]{Ignacio Maga\~na~Hernandez}
\affiliation{\CMU}
\affiliation{\Milwaukee}

\author[0000-0001-6145-8187]{Virginia D’Emilio}
\affiliation{\Cardiff}

\author[0000-0002-8445-6747]{Soichiro Morisaki}
\affiliation{\UTokyo}

\begin{abstract}
Fast radio bursts (FRBs) are a newly discovered class of radio transients that emerge from cosmological sources and last for $\sim$ a few milliseconds. However, their origin remains a highly debated topic in astronomy. 
%Among cataclysmic events, binary neutron star (BNS) mergers are proposed to be promising multi-messanger counterparts of at least some fraction of FRBs. However, these associations should not be made solely on the basis of low chance association probability. 
Among the plethora of cataclysmic events proposed as potential progenitors, binary neutron star (BNS) mergers have risen as compelling candidates for at least some subset of apparently non-repeating FRBs. However, this connection should not be drawn solely on the basis of chance coincidence probability.
In this study, we delineate necessary criteria that must be met when considering an association between FRBs and BNS mergers, focusing on the post-merger ejecta environment. 
%we present necessary conditions pertaining to the post-merger ejecta environment that must be considered before associating FRBs with BNS mergers. 
 To underscore the significance of these criteria, we scrutinize the proposed association between \GW and \FRB. Our investigation meticulously accounts for the challenging condition that the FRB signal must traverse the dense merger ejecta without significant attenuation to remain detectable at 400 MHz.
%To illustrate the significance of these criteria, we examine the proposed association of \GW and \FRB.
%, we consider the example of 
%\GW and \FRB. Recent studies have argued for a possible association between the binary neutron star (BNS) merger \GW and \FRB at a confidence level of 2.8$\sigma$. 
%The authors argue that the observations are consistent with a long-lived highly magnetized supramassive neutron star (SMNS) that formed after the BNS merger and was stable for approximately 2.5 hours before promptly collapsing into a black hole. 
%We investigate this proposed association, carefully considering the constraint that the FRB signal must traverse the high-density merger ejecta without experiencing noticeable attenuation to enable its detection at 400 MHz. 
Furthermore, we find that if the FRB is indeed linked to the gravitational wave event, the GW data strongly support a highly off-axis configuration, with a probability of the BNS merger viewing angle $p(\theta_v$ $>$ 30$^{\circ}$) to be $\approx 99.99$\%. Our findings therefore strongly exclude an on-axis system, which we find on the other hand to be required in order for this FRB to be detectable. Hence, we conclude that \GW is not related to \FRB. We also discuss implications of our results for future detections of coincident multi-messenger observations of FRBs from BNS remnants and GW events and argue that BNS merger remnants cannot account for the formation of $>$ 1\% of FRB sources. 
This observation suggests that short gamma-ray bursts should not be used to explain global attributes of the FRB host population.
%An immediate consequence of this conclusion is the reconsideration of short gamma-ray bursts as viable explanations for the overarching characteristics of the FRB host population.
%An immediate implication of this result is that short GRBs should not be considered to explain the global properties of FRB host population.
\end{abstract}


\section{Introduction} \label{section:introduction}

% GW170817, coincident high significance

Fast radio bursts (FRBs) are intense flashes of highly coherent radio waves (brightness temperature $\sim 10^{36}$ K) that exhibit duration ranging from microseconds to milliseconds \citep{Lorimer2007}. These bursts can be detected across vast extragalactic distances making them a promising cosmological probe \citep{Cordes2019,Caleb2021,Petroff2022}. Since the discovery of the first FRB in 2007, over 1000 FRBs have been reported in the scientific literature\footnote{For a complete list of known FRBs, see \url{https://www.herta-experiment.org/frbstats/} or the TNS \citep{2020TNSAN..70....1Y}.}. Nevertheless, their origins remain a topic of continued scientific investigation and speculation. Recent years have seen significant advancements in understanding the underlying source population of these cosmic events. For example, \cite{2021ApJS..257...59C} inferred a sky rate of bright FRBs (bursts of fluence $\geq$ 5 Jy ms at 600 MHz) to be $\sim$ 600 FRBs/sky/day, and estimated the volumetric rate of FRBs with a fiducial energy of $\sim 10^{39}$ ergs to be $\sim 10^{5}$ FRBs/yr/Gpc$^{3}$ \citep{2023ApJ...944..105S}. Moreover, there exists a substantial sub-population of FRBs that repeats \citep{2016Natur.531..202S,2019Natur.566..235C,2019ApJ...885L..24C,2020ApJ...891L...6F,2021MNRAS.500.2525K,2021ApJ...910L..18B,2022ATel15679....1M,2022Natur.606..873N,2022ApJ...927...59L,2023ApJ...947...83A}. These observations suggest that most of the FRBs, particularly those in the nearby Universe, are repeaters, suggesting non-cataclysmic origins \citep{2021ApJ...919L..24B} and are likely have progenitors formed via core-collapse supernovae \citep{2023arXiv231010018B}. However, cataclysmic origins for a subset of FRBs that do not exhibit repetition cannot be ruled out.

Several FRB source models invoke cataclysmic events involving the coalescence of compact binary systems consisting of neutron stars and/or black holes, to explain a fraction of non-repeating FRBs \citep{2013PASJ...65L..12T,2014A&A...562A.137F,2015ApJ...814L..20M,2016ApJ...822L...7W,2016ApJ...827L..31Z,2018ApJ...868...17A}. It is worth noting that many of these events in the local Universe ($\sim$ 100 Mpc) can be detected using current-generation gravitational wave detectors 
\citep{,2021ApJ...915...86A,2022arXiv220312038T}. Binary neutron star (BNS) mergers, within the realm of cataclysmic events, are regarded as potential sources of FRB-like signals at different stages of the merger: prior to, during, or after the merger \citep[the full list of models, see][]{2019PhR...821....1P}. Recently, \cite{moroianu2022assessment} reported a possible association, at the 2.8$\sigma$ level, between \GW \citep{LIGOScientific:2020aai}, the second BNS merger to be detected in gravitational waves, and \FRB \citep{2021ApJS..257...59C,2023arXiv231010018B}. \GW was detected by the LIGO-Virgo-KAGRA (LVK) gravitational wave detector network during the first half of their third observing run, O3a \citep{LIGOScientific:2021usb}. The immediately interesting aspect of this merger was that its chirp ($\sim 1.44~M_\odot$) and total mass ($\sim 3.4~M_\odot$) were significantly larger than any other known BNS system. Because one of the LIGO detectors was not in observing mode at the time of detection, the sky localization of this event is not precise (90\% credible localization region is $7461~\rm{deg}^2$). Moreover, since it was at a significantly larger distance (luminosity distance $d_L=159^{+69}_{-71}$ Mpc) compared to GW170817 ($\sim 40$ Mpc), any associated electromagnetic counterpart would have been significantly fainter, and the large localization volume challenged both follow-up searches and high-confidence association for any candidate counterpart. No significant association was therefore found between the GW alert and plausible counterparts.

\FRB was discovered by the Canadian Hydrogen Intensity Mapping Experiment/Fast Radio Burst \citep[CHIME/FRB;][]{2021ApJS..257...59C} project $\approx 2.5$ hours after the GW alert, and according to \cite{panther2022most}, it most likely occurred in the host galaxy \GAL (probability of true association $\approx$ 0.79. This association was found to be true by \cite{2023arXiv231010018B} using CHIME/FRB baseband localization (chance association probability $<$ 0.1\%). Since the total mass for \GW was found to be $3.4^{+0.3}_{-0.1} \ M_{\odot}$, the LVK collaboration suggested that the merger of the two neutron stars, most likely, promptly collapsed into a black hole, given our current constraints on the equation of state (EOS) for dense nuclear matter \citep{abbott2020gw190425}. To explain the discrepancy between the LVK interpretation and delay of 2.5 hours post-merger, \cite{moroianu2022assessment} and \cite{zhang2022physics} put forward a proposed scenario for \GW that could have led to the formation of a highly spinning supramassive neutron star with an extremely strong magnetic field ($\geq 10^{14}$ Gauss) and the EOS consistent with proposed quark models. In this context, two primary inquiries arise: Firstly, can we anticipate the occurrence of Fast Radio Bursts (FRBs) within such a system? Secondly, how might we employ empirical evidence exclusively to substantiate or challenge these potential associations?
% \fix{In order for the proposed NS to not collapse directly into a black hole, this might be due to increased mass support due to the required highly spinning remnant as well as the necessity for a stiffer EOS and potentially an exotic compact object as one of the binary components, e.g., a quark star.}
%In order for the proposed neutron star to not collapse directly into a black hole, it is necessary to invoke a highly spinning remnant, as this might provide increased mass support, as well as a stiffer EOS and potentially an exotic compact object as one of the binary components, e.g., a quark star. The hypermassive NS would then survive the direct collapse for about 2.5 hours until it collapses into a black hole ejecting its magnetosphere in the process leading to the production of \FRB. 

In this work, we present necessary conditions based on astrophysical and GW constraints that any plausible FRB and BNS merger association must satisfy before even assessing the probability of chance coincidence of the two events. This is because the FRB signal must traverse the high-density merger ejecta without experiencing noticeable attenuation to enable its detection at $\sim$ 1 GHz. To demonstrate the use of the aforementioned constraint, we reconsider the association between \GW and \FRB using two classes of constraints that were not discussed in \cite{moroianu2022assessment}: (1) An updated GW parameter estimation under the assumption that \GAL is the host galaxy of both \FRB and \GW %\citep{gw190425:association} 
which is discussed in \S\ref{section:pe}, and (2) the effect of the BNS merger ejecta in the propagation of the FRB signal which is discussed in \S\ref{sec:propagation}. 
%We stress here that the association is not claimed to be conclusive by \cite{moroianu2022assessment}, but rather marginal since it is $<3\sigma$. 
%The purpose of this work is to explore whether astrophysical and GW constraints would support or disfavor this possible association.
%With these constraints, we examine the association taking into account the consistency of the measured viewing angle with the blitzar model invoked in \citet{moroianu2022assessment}, the lack of a short Gamma-Ray Burst (GRB) and kilonova detection, and also determined the ejecta mass that can facilitate the propagation of the FRB signal at 400 MHz without any noticeable attenuation.
Considering these new constraints, we examine the association by evaluating the consistency of the measured viewing angle with the blitzar model invoked in \cite{moroianu2022assessment}. 
%Additionally, we take into account the absence of a short gamma-ray burst (GRB) and kilonova detection. 
Furthermore, we determine the maximum ejecta mass required to enable the propagation of the FRB signal at 400 MHz without experiencing any noticeable attenuation.
Both of these constrained independently disfavor the association of \GW and \FRB with high confidence as presented in \ref{section:related}. Finally, we discuss the implications of our findings for future associations between FRBs and BNS merger remnants in \S\ref{section:frb-model} and conclude in \S\ref{section:conclusion}.
%calculate the probability of prompt collapse as in \citep{agathos2020inferring,abbott2020gw190425}. 

\section{Parameter estimation with \GAL as the host galaxy} \label{section:pe}

\begin{figure}[ht!]
    \centering
    \includegraphics[width=0.5\textwidth]{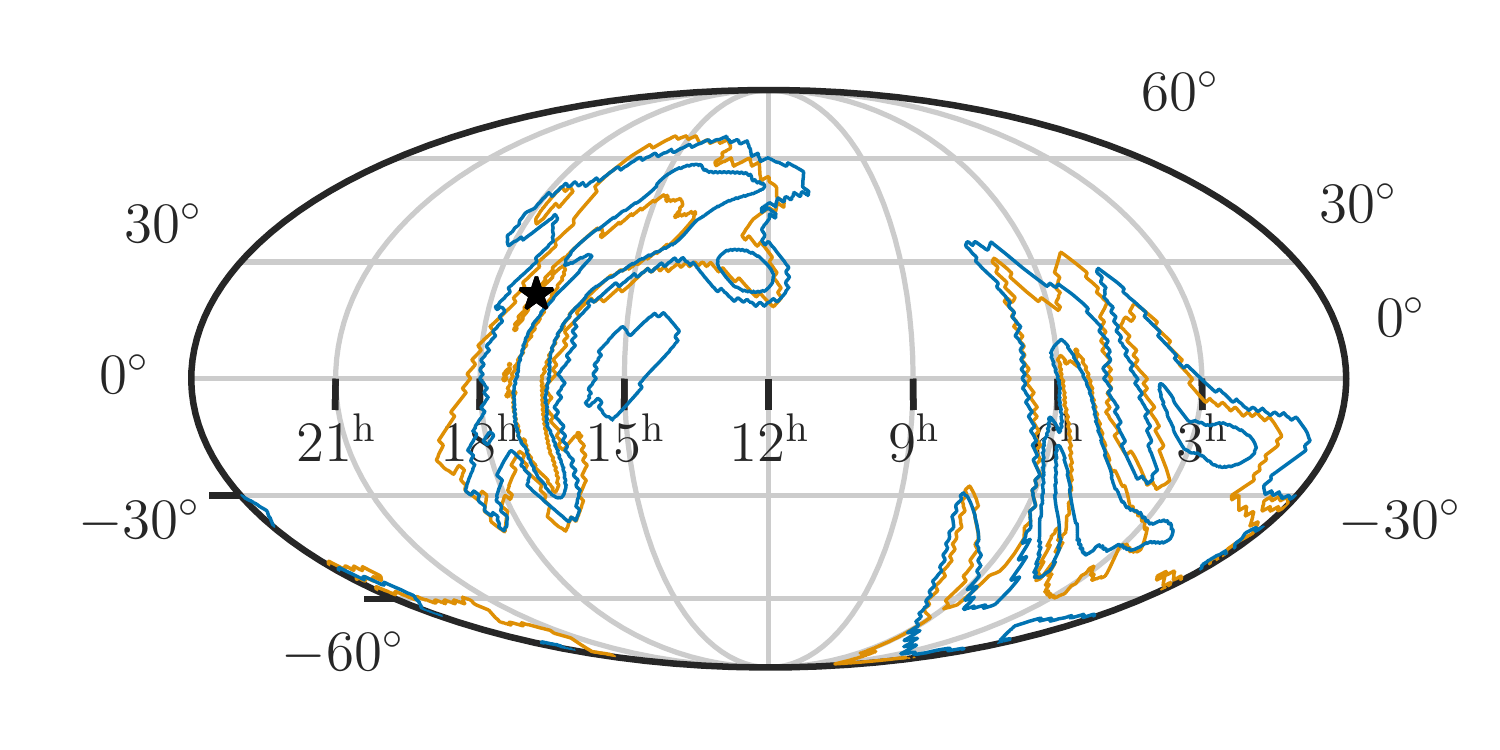}
    \caption{Sky localization regions using both the low spin (blue) and high spin (yellow) posterior samples as reported in \cite{abbott2020gw190425}. We show both the 50\% and 90\% confidence intervals for each case and show the location of \GAL. }\label{figure:skymap}
\end{figure}

In order to assess the viability of \GAL as the host for \GW and \FRB, we use the parameter estimation results of \cite{gw190425:association}. Here, we present a succinct overview of the Bayesian parameter estimation procedure employed in this study. We use the \texttt{BILBY} library \citep{Ashton:2018jfp, Romero-Shaw:2020owr} and the \texttt{DYNESTY} nested sampler package \citep{Speagle:2019ivv}. The waveform model used in the analysis is \texttt{IMRPhenomPv2\_NRTidal} \citep{Dietrich:2017aum, Dietrich:2018uni}, which includes both precession and tidal effects. To make the analysis computationally inexpensive, we use the reduced order quadrature technique \citep{Smith:2016qas, amanda_baylor_2019_3478659}.% to reduce the computational cost of the waveform evaluations. 

Following the LVK analysis conventions presented in \cite{LIGOScientific:2021usb}, we perform two sets of analyses with different spin priors, namely, a \low and a \high prior where we assume uniform distributions on the dimensionless spin magnitudes for both components to be within the ranges $\chi<0.05$ and $\chi<0.89$, respectively. The prior probability distributions on the remaining binary parameters used in this work are the same as that used in \cite{LIGOScientific:2021usb}, except that: 1) we fix the sky position, i.e., we set the sky location to the location of \GAL $(\alpha, \delta)=(255.72^\circ, 21.52^\circ)$ from \cite{2020ApJS..249....3A}, and 2) we additionally fix the redshift to that of \GAL \citep[spectroscopic redshift, $z =  0.03136 \pm 0.00009$;][]{SDSS:2008tqn}. It is worth noting that changing the position of the BNS merger within the 90\% confidence baseband localization region of the FRB estimated by \cite{2023arXiv231010018B} does not have any meaningful impact on our results. Finally, for completeness, we show the sky localization posteriors and the location of \GAL in Figure \ref{figure:skymap}. We note that the location of the galaxy is within the 90\% confidence region under both low and high spin assumptions. 

We then calculate the viewing angle, $\theta_v = \min(\theta_{\rm{JN}}, 180^\circ - \theta_{\rm{JN}})$, using the measured inclination angles $\theta_{\rm{JN}}$ for each of the cases considered. In particular, we note that we can measure the viewing angle to \GW under the fixed position assumption since the measured redshift to \GAL breaks the distance-inclination degeneracy \citep{2011CQGra..28l5023S}. We also compute the total mass, $m_{\rm{total}}$ in the source frame for each case.  We report our viewing angle and mass measurements in Table \ref{table:parameters} using the \low and \high prior results with both the fixed sky and fixed position assumptions. Additionally, we show the marginalized posteriors on $\theta_v$ for the low and high spin prior cases in Figure \ref{figure:mass_viewing}. Finally, the updated mass measurements are used to calculate an up-to-date estimate for the probability of \GW promptly collapsing into a black hole following \cite{abbott2020gw190425,agathos2020inferring} in \S\ref{section:discussion}.
%and show in Figure \ref{figure:mth}.

\begin{deluxetable*}{lcccc} \label{table:parameters}
\tablecaption{Summary of the updated mass parameters and viewing angle for \GW using both the \low and \high priors under the fixed sky and fixed position assumptions as described in \S\ref{section:pe}. We report all mass measurements in the source frame assuming a Planck 2015 cosmological model.
}
\tablehead{
    \colhead{~} & \multicolumn{2}{c}{Low Spin Prior} & \multicolumn{2}{c}{High Spin Prior}\\
    \colhead{~} & \colhead{Fixed Sky} & \colhead{Fixed Position} & \colhead{Fixed Sky} & \colhead{Fixed Position}
}
\startdata
Primary mass $m_1/M_{\odot}$ & $1.74^{+0.17}_{-0.09}$ & $1.75^{+0.17}_{-0.09}$ & $2.01^{+0.53}_{-0.33}$ & $2.10^{+0.59}_{-0.40}$\\
Secondary mass $m_2/M_{\odot}$ & $1.55^{+0.08}_{-0.14}$ & $1.57^{+0.08}_{-0.13}$ & $1.35^{+0.26}_{-0.25}$ & $1.32^{+0.30}_{-0.26}$\\
Chirp mass $\mathcal{M}/M_{\odot}$ & $1.43^{+0.02}_{-0.02}$ & $1.442^{+0.001}_{-0.001}$ & $1.43^{+0.02}_{-0.02}$ & $1.442^{+0.001}_{-0.001}$\\
Mass ratio $m_2/m_1$ & $0.89^{+0.10}_{-0.15}$ & $0.89^{+0.10}_{-0.15}$ & $0.67^{+0.29}_{-0.24}$ & $0.63^{+0.32}_{-0.24}$\\
Total mass $m_\mathrm{tot}/M_{\odot}$ & $3.30^{+0.06}_{-0.04}$ & $3.32^{+0.04}_{-0.01}$ & $3.37^{+0.28}_{-0.11}$ & $3.42^{+0.34}_{-0.11}$\\
%%
%Effective inspiral spin parameter $\chi_\mathrm{eff}$ & $0.01^{+0.02}_{-0.01}$ & $0.01^{+0.02}_{-0.01}$ & $0.06^{+0.08}_{-0.05}$ & $0.07^{+0.10}_{-0.06}$\\
%%
%Luminosity distance $D_\mathrm{L}$ & $183.7^{+58.2}_{-75.3} \ \rm{Mpc}$ & $-$ & $183.2^{+57.8}_{-73.3} \ \rm{Mpc}$ & $-$\\
%%
Viewing angle $\theta_v$ & $37.8^{+42.4}_{-27.5} \ \rm{deg}$ & $56.1^{+14.3}_{-9.7} \ \rm{deg}$ & $37.8^{+41.3}_{-26.9} \ \rm{deg}$ & $55.6^{+14.3}_{-9.2} \ \rm{deg}$\\
\enddata
\end{deluxetable*}

\section{Characterizing Propagation Effects on 400 MHz FRB Emission}
\label{sec:propagation}

In \cite{2023NatAs.tmp...63M}, the authors employed the ``blitzar" mechanism first proposed by \cite{2014A&A...562A.137F} to provide an explanation for FRB 20190425A at time t$_{\rm FRB}$ = 2.5 hours after the GW190425 event. In doing so, the authors made two assumptions: firstly, that GW190425 represents a BNS merger event, and secondly, that the resulting post-merger remnant is a supramassive neutron star (SMNS). We note that the validity of both assumptions is still being strongly debated \citep{2020ApJ...891L...5H,2020ApJ...892L...3A,2020MNRAS.494..190F,2020ApJ...904...80E,2021A&A...654A..12B}. In the blitzar mechanism, when a highly magnetized SMNS collapses to form a black hole, it ejects its magnetosphere as the magnetic fields cannot puncture the event horizon, as dictated by the no-hair theorem \citep{2015PhRvL.114o1102G}. This generates strong magnetic shock waves that accelerate electron and positron pairs already present in the magnetosphere to relativistic velocities. The FRB is produced via curvature radiation emission when these relativistic particles follow the distorted ejected magnetic field lines over spatial scales $\approx$ c$\Delta$t, where $\Delta$t is the temporal width of the FRB. However, we note that is it still unclear if the curvature radiation can be operational in such a chaotic scenario \citep{2017MNRAS.468.2726K}. We ignore these concerns in the following discussions and assess the feasibility of this scenario for this specific GW event. 

In order to estimate the ejecta electron number density, n$_{\rm ej}$, we assume spherically symmetric and homogeneous ejecta of mass M$_{\rm ej}$, which expands with velocity v$_{\rm ej}$ = 0.2c, where c is the speed of light and has the electron fraction $\rm Y_{e} = 0.2Y_{e,0.2}$. The ejecta is fully ionized\footnote{For GW190425, this assumption is challenged by \cite{2023arXiv230915195R} based on their numerical relativity simulation results. However, this won't affect the major conclusion from this study.} and composed of heavy elements with Z/A $\approx$ 2. Our choice of these fiducial values is informed by both recent simulation results and observations from GW170817 \citep{2013PhRvD..87b4001H,2017ApJ...850L..37P,2018MNRAS.480.3871C}. Note that the spherical symmetry of the early epoch ejecta appears reasonable for various analytical estimates \citep{2019LRR....23....1M} and has been recently confirmed for AT2017gfo, the kilonova associated with the GW170817 gravitational wave event \citep{2023Natur.614..436S}. However, we discuss the non-spherical scenario in \S\ref{section:related}.

In the spherically symmetric ejecta scenario, we estimate n$_{\rm ej}$ to be,
 
\begin{equation}\label{Eq:ne}
 \rm n_{ej} \approx \frac{3M_{ej}Y_{e}}{8\pi m_{p}(v_{ ej}t)^{3}} = 1.8 \times 10^{14} \frac{M_{ej}}{M_{\odot}} Y_{e,0.2}v_{ej,0.2c} cm^{-3},    
\end{equation}

where m$_{\rm p}$ is the proton mass. 
Note that in all of our calculations, we fix t = t$_{\rm FRB}$ = 2.5 hours.

In order for us to observe FRB 20190425A, which exhibits a flat and broadband frequency spectrum \citep[the FRB is detected in the full CHIME band, i.e., 400 - 800 MHz,][]{2021ApJS..257...59C}, the ambient environment must be sufficiently rarefied, and hence, should be optically thin to various propagation effects that can suppress the FRB signal at 400 MHz. Using this constraint, we now derive an upper limit on the ejecta mass of GW190425 for different propagation effects described below. 

%When two neutron stars merge, they release a large amount of energy in the form of gravitational waves, and also eject a significant amount of material into the surrounding environment. This material, known as the binary neutron star ejecta, is a hot and dense mixture of neutron-rich material, and can give rise to a range of electromagnetic signals such as kilonovae and gamma-ray bursts.

\subsection{Plasma absorption constraint}

The plasma frequency v$_{\rm p}$ of electrons in the BNS ejecta must be smaller than 400 MHz, otherwise, the FRB signal would have been absorbed. Therefore,  v$_{\rm p}$ given by, 
\begin{equation}\label{Eq:plasma_abs}
\rm \rm v_{p} = \frac{1}{2\pi}\sqrt{\frac{n_{ej}e^{2}}{m_{e}\varepsilon_{0}}} < 400 ~MHz,
\end{equation}

where m$_{\rm e}$ is the electron mass, $\varepsilon_{0}$ is the permittivity of free space, and e is the electron charge. Using Equation \ref{Eq:ne} in Equation \ref{Eq:plasma_abs}, we get M$_{\rm ej} < 1 \times 10^{-5} M_{\odot}$.

\subsection{FRB dispersion measure constraint}
\noindent  \cite{2021ApJS..257...59C} reported the dispersion measure (DM) of FRB 20190425A, defined as 
\begin{equation}\label{eq:dm}
    \rm DM = \int_{0}^{L} n_{e} \,dx,
\end{equation} where n$_{e}$ is the free electron number density and L is the comoving distance of the FRB source from us, to be 128.1 pc cm$^{-3}$. Note that there are several astrophysical components that contribute to the FRB DM, namely, the Milky Way interstellar medium (MW ISM) and its halo, the intergalactic medium (IGM), and the FRB host and its local environment (in this case, the local environment would be the BNS ejecta), such that $\rm DM_{FRB} = DM_{MW ISM} + DM_{MW halo} + DM_{IGM} + DM_{host} + DM_{BNS ejecta} = 128.1~pc~cm^{-3}$.
Here, DM$_{\rm IGM} \sim$ 30 pc cm$^{-3}$ using the Macquart relation \citep{macquart2020census}, if we assume UGC 10667 at redshift =  0.0312 as the FRB host, DM$_{\rm MW ISM} \sim 40$ pc cm$^{-3}$ along the FRB line-of-sight using the NE2001 \citep{cordes2002ne2001} and YMW16 \citep{yao2017new} Galactic electron density distribution models, and DM$_{\rm MW halo} \sim 30$ pc cm$^{-3}$ from the \cite{yamasaki2020galactic} Milky Way Halo model. Using these values, we estimate $\rm DM_{host} + DM_{BNS ejecta} \sim 30~pc~cm^{-3}$ and compute an upper limit on the ejecta mass assuming $\rm DM_{host} = 0~pc~cm^{-3}$ using Equations \ref{Eq:ne} and \ref{eq:dm} to be $\lesssim$ 10$^{-8} M_{\odot}$. Note that uncertainties in the DM estimates of different astrophysical components would result in $<$ 1 dex. error in our upper limit estimate.

\subsection{Induced Compton scattering constraint}

Induced Compton scattering, also known as stimulated Compton scattering, is a process in which a photon collides with a free-charged particle, such as an electron, and transfers a portion of its energy to the particle \citep{1982MNRAS.200..881W}. As a result of the collision, the photon is scattered and its energy is reduced, while the electron gains kinetic energy. In the case of short (duration = $\rm \Delta t \sim$ 1 ms) coherent radio emission, like FRB, with brightness temperature $\rm T_{B}~\sim~10^{36}$ K traversing a non-relativistic plasma, the induced Compton scattering optical depth can be estimated using Equations 6 and 14 from \cite{2008ApJ...682.1443L},
\begin{equation}\label{eq:ICS}
    \rm \tau_{C} = \rm \frac{3\sigma_{T} c K_{B}}{8 \pi m_{e}} n_{ej} T_{B} \frac{{\Delta t}^{3}}{{v_{ej}t}^{2}} < 1, 
\end{equation}
%\approx 10^{-10} T_{B,36}{v_{ej,0.2c}^{-2} \Delta T_{1}^{3}
where $\rm \sigma_{T}$ is the Thomson cross-section of electrons. Equation \ref{eq:ICS} gives $ \rm M_{ej} \lesssim 5 \times 10^{-11} M_{\odot}$.

\subsection{Simulated Raman scattering constraint}

In the case of highly coherent sources, like FRBs, the energy density of the radio waves is so high that it can create Langmuir waves in dense ejecta plasma through a process called parametric decay instability \citep{1995MNRAS.274..717L}. The Langmuir waves can then interact with FRB emission and induce Raman scattering, in which the photons are scattered away and the FRB signal decays away to a lower frequency and momentum state. The effect of simulated Raman scattering is significant in our case when

\begin{equation}
    \rm 400 \gtrsim 130 \left(\frac{n_{e}}{cm^{-3}}\right)^{1/2} \left(\frac{T_{ej}}{K}\right)^{-1/2} MHz.
\end{equation}

Figure \ref{figure:ejecta_mass} shows the maximum BNS ejecta mass as a function of the ejecta temperature. At the fiducial ejecta temperature T$_{\rm ej}$ of kilonova emission of $\approx 10^{4}$ K \citep{2019LRR....23....1M}, we estimate $ \rm M_{ej} \lesssim 5 \times 10^{-10} M_{\odot}$.

\subsection{Razin suppression constraint}

The Razin suppression \citep{razin1960theory}, also known as the Razin-Tsytovich effect \citep{1972Ap&SS..18..267M}, is a non-linear plasma effect in the presence of a strong magnetic field that can result in the attenuation of radio emission at $\nu \gtrsim$ v$_{\rm p}$. In our case, it can suppress the production of bunched coherent curvature emission \citep{1986ApJ...302..120A} at a frequency below the Razin cut-off frequency ($ \rm \nu_{R}$). We use the following relation from \cite{2019ApJ...874...72R},
\begin{equation}
  \rm \rm \nu_{R} = 400 \left(\frac{n_{ej}}{cm^{-3}}\right)^{1/2}\left(\frac{T_{ej}}{K}\right)^{-1/2} < 400~MHz,  
\end{equation}

and estimate $ \rm M_{ej} \lesssim 6 \times 10^{-11} M_{\odot}$ at T$_{\rm ej}$ = 10$^{4}$ K. At other temperatures, the maximum $\rm M_{ej}$ constraint is shown in Figure \ref{figure:ejecta_mass}.

\subsection{Free-free absorption constraint}

\begin{figure*}[ht!]
    \centering
    \includegraphics[width=0.99\textwidth]{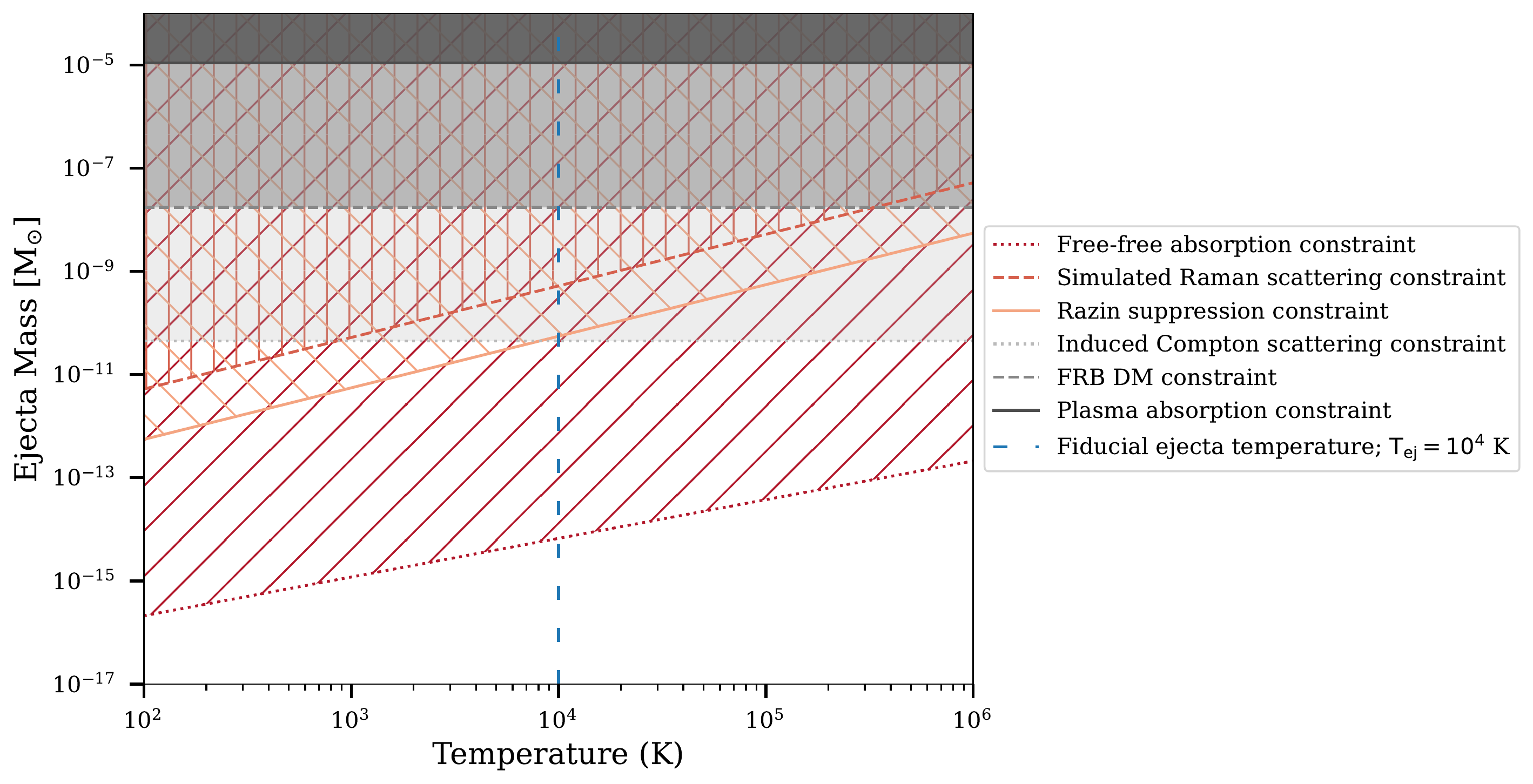}
    \caption{Derived constraints on the BNS ejecta mass (M$_{\rm ej}$) if FRB 20190425A is indeed associated with GW190425. Hatched regions are excluded using ejecta temperature (T$_{\rm ej}$) dependent propagation effects and shaded regions are excluded due to constraints which are independent of the ejecta temperature as discussed in \S\ref{sec:propagation}.}\label{figure:ejecta_mass}
\end{figure*}

An FRB signal can interact with the free ions and electrons in the surrounding ejecta medium through free-free absorption which shows a characteristic frequency dependence $\sim \nu^{2}$ when the free-free optical depth  ($\tau_{\rm ff}$) $>$ 1. Note that the amount of absorption of the FRB signal depends on both the electron number density and temperature (T$_{\rm ej}$) of the ejecta. In our case, as the FRB spectrum does not exhibit such behaviour, $\tau_{\rm ff}$  must be $<$ 1 at 400 MHz. At T$_{\rm ej} \approx 10^{4}$ K, we estimate an upper limit on M$_{\rm ej}$ at $\nu = 0.4$ GHz using Equation 35 from \cite{2020ApJ...891...72W}, i.e.,
\begin{equation}
\rm \tau_{ff} \simeq 2.3 \times 10^{28}  Y_{e,0.2}^{2} v_{ej,0.2c}^{-5} \nu_{0.4}^{-2} T_{ej,4}^{-3/2} M_{ej,\odot}^{2},    
\end{equation}

 to be 7 $\times 10^{-15} M_{\odot}$. This is the most stringent constraint on the  M$_{\rm ej}$. The M$_{\rm ej}$ estimate at other  T$_{\rm ej}$ are shown in Figure \ref{figure:ejecta_mass}. We note that changing other kilonova ejecta parameters within reasonable ranges will not affect this estimate by more than an order of magnitude, so the constraint remains very stringent.
 
\subsection{Synchroton self-absorption constraint}

If the post-merger remnant is a supramassive neutron star, it is expected to have a strong magnetic field of at least $10^{14}$ G \citep{2015ApJ...809...39G}, which could create a pulsar wind nebula after interacting with the ejecta. This nebula would accelerate charged particles to relativistic speeds and hinder the transmission of the FRB signal via synchrotron self-absorption. Hydrodynamic simulations by \cite{2018PASJ...70...39Y} suggest that the dynamical ejecta would obstruct the escape of coherent radio emission until $\sim$ 1-10 years after the merger. However, quantifying the effect of ejecta mass on the synchrotron self-absorption optical depth is a complex problem that depends on various parameters, such as the electron energy spectrum. Further discussion on this will be presented elsewhere.

\section{Discussion} \label{section:discussion}

\subsection{Are \FRB and \GW related?} \label{section:related}

In \S\ref{sec:propagation}, we showed that if the FRB is indeed associated with GW190425, then for the FRB emission to pass through the BNS ejecta without suffering significant attenuation due to different propagation effects, the ejecta mass has to be $\lesssim 10^{-15}$ M$_{\odot}$. This is orders of magnitude smaller than what has been observed in the case of GW170817 and predicted by numerical simulations \citep[$\sim 10^{-2} - 10^{-4}$ M$_{\odot}$;][]{2013PhRvD..87b4001H,2015PhRvD..91l4041D,2016MNRAS.460.3255R,2017ApJ...850L..39A,2022PhRvD.106h4039D,2023PhRvD.107f3028H}.

The only possible scenario in which the FRB emission can escape through the BNS ejecta is if the outflow electron number density is extremely low ($\sim$ 0.1 cm$^{-3}$ using M$_{\rm ej} = 10^{-15}$ M$_{\odot}$ in Equation \ref{Eq:ne}). To explain this, \cite{2014ApJ...780L..21Z} proposed a scenario in which the FRB is emitted in the direction of the relativistic jet generated after the merger of binary neutron stars, which is observed as a short-duration Gamma-Ray Burst (GRB), where the jet's action can lead to a substantial reduction in the surrounding ejecta density. In this case, we expect the inclination angle of the merger to be $<$ 30$^{\circ}$ \citep{2016MNRAS.460.3255R,2018ApJ...857..128J,2020LRR....23....4B}. However, this scenario can be disfavoured if we believe the FRB-GW association. As shown in Table \ref{table:parameters} and Figure \ref{figure:ejecta_mass}, if the FRB is associated with UGC 10667, we expect the BNS merger viewing angle to be $\theta_v \sim 60$ deg, making the GW event an off-axis merger. In fact, our posteriors do exclude an on-axis system, and an off-axis system similar to GW170817 at $p(\theta_v>30^o)=99.99$\%. We note that this conclusion does not change whether we assume a low or high spin prior. This consideration also rules out a possible GW-FRB association with the weak GRB detection by the Anti-Coincidence Shield (ACS) on INTEGRAL \citep{Pozanenko}, as opposed to the consistency with the GW-FRB association, reported in \citet{moroianu2022assessment}. Finally, we note that it is not clear if the polar region remains low-density after $\sim$ 2.5-hrs the launch of the relativistic jet \citep{2019ApJ...886..110M,2019LRR....23....1M,2023Natur.614..436S}. More importantly, MHD simulations suggest that forming a relativistic jet when the BNS merger remnant is a neutron star is more challenging than when it is a black hole \citep{2020ApJ...900L..35C,2021MNRAS.500..627H}. 
%Therefore, we conclude that the FRB 20190425A is not associated with GW190425.
%On the other hand, it is worth noting that the non-detection reported for Fermi \citep{song_190425} is inconclusive for this work as the FRB location was in the Earth occultation zone for Fermi at the time of the GW event. 

%The lack \citep{song_190425} of a high confidence Gamma-Ray Burst (GRB) detection associated with this event is consistent with our PE results showing an off-axis ($\theta_v \sim 60$ deg) binary system at the location of the FRB, assuming the GW-FRB association.

\begin{figure*}[ht!]
    \centering
    \includegraphics[width=0.45\textwidth]{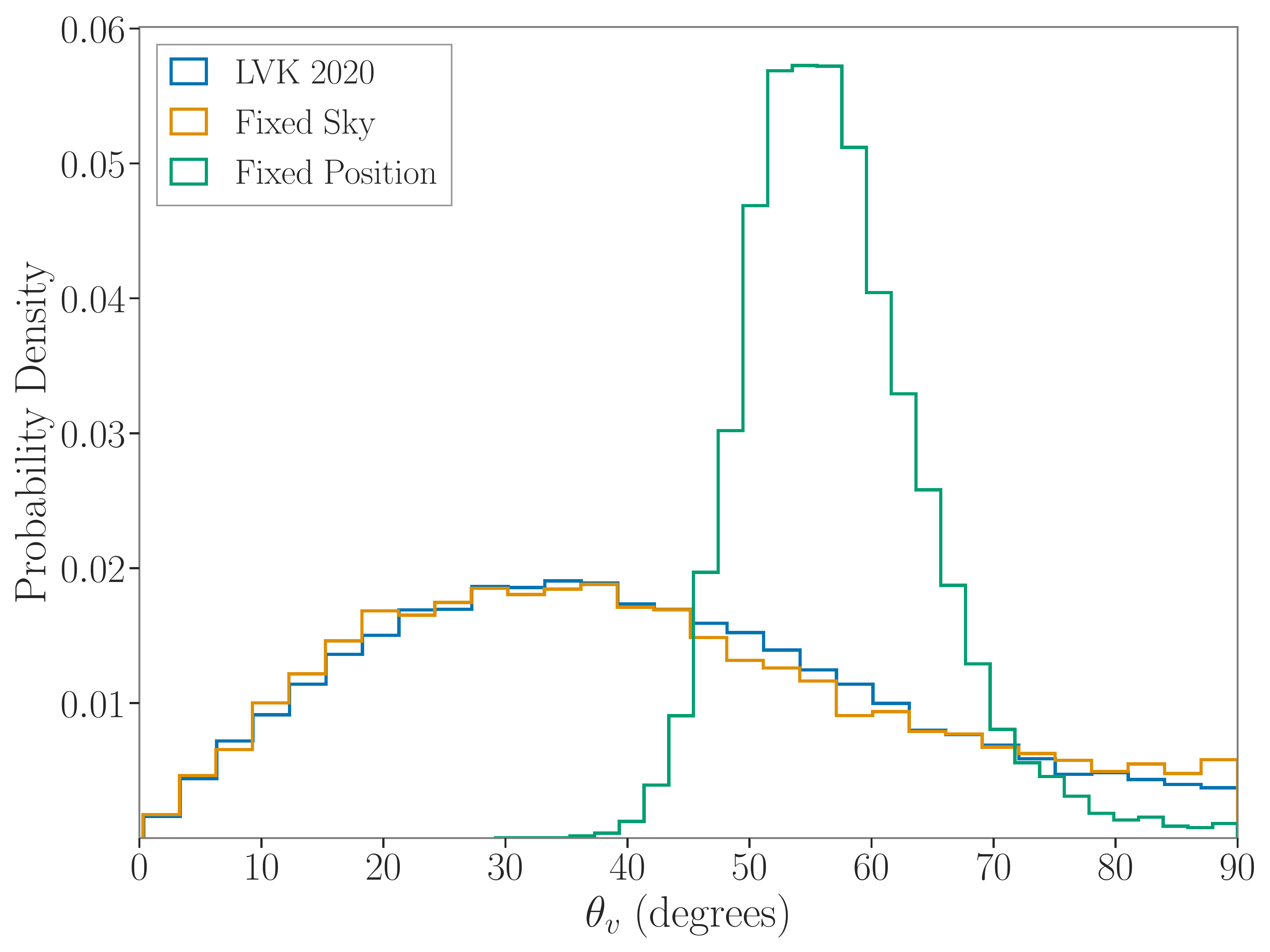}
    \includegraphics[width=0.45\textwidth]{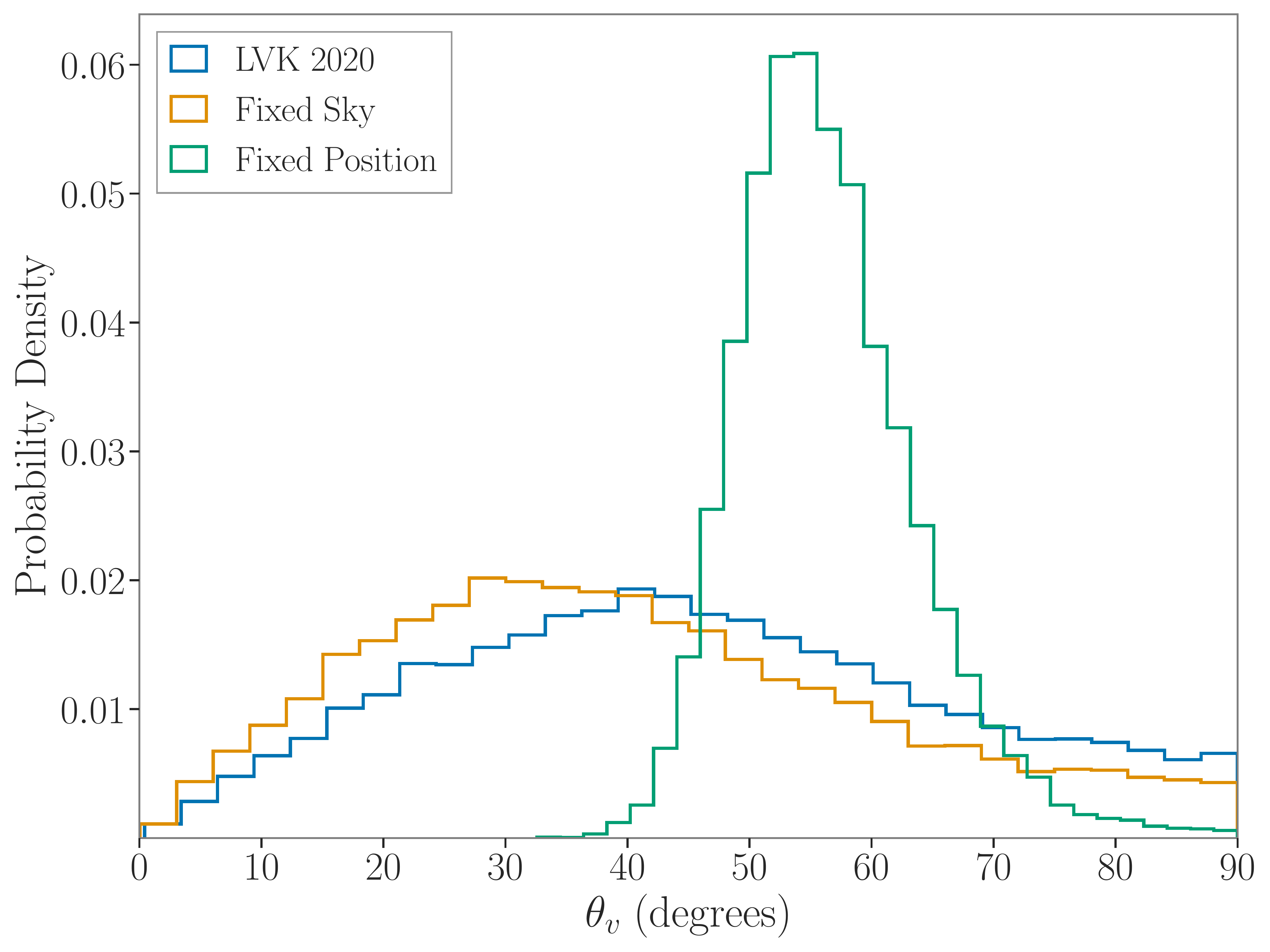}
    \caption{Posterior distributions on the viewing angle $\theta_v$ for \GW for both the low spin (left panel) and high spin (right panel) priors under the fixed sky location and fixed position assumptions as described in \S\ref{section:pe}. For both cases, we also show the viewing angle posteriors computed using the results of \cite{abbott2020gw190425} for reference. }\label{figure:mass_viewing}
\end{figure*}

%The authors noted that  for an FRB to be observed following the merger, the line of sight needs to be cleared by a sGRB jet. 
We note that our electron density calculations rely on the assumption that the BNS ejecta is isotropic, whereas in reality some degree of anisotropy is expected (e.g. \citealt{darbha}).
%, where simulations suggest otherwise (e.g. Rosswog et al. 1999) where the most of the matter is ejected close to the equatorial plane (within 60 degrees). 
However, in most general-relativistic simulations of binary neutron star mergers, the ejecta is found to be mostly distributed over a broad $\sim$ 60$^{\circ}$ angle from the equatorial plane \citep{2016MNRAS.460.3255R}, with the addition of  squeezed dynamical ejecta possible along the polar region (e.g. \citealt{Kasen_2017}). This would not affect our conclusion because if the GW event were associated with the FRB, the GW data suggests 
that the inclination angle (see Figure \ref{figure:mass_viewing}) would still be within a region where the ejecta are expected. Moreover, our constraints only consider dynamical ejecta which is composed of a combination of tidally- and shock-driven ejecta, and ignore the contribution from near-isotropic remnant wind \citep{2013ApJ...771...86G} and accretion disk, which consist of the material expelled from the central part of the remnant because of tidal torques or from the collision interface settles into a thick accretion disk \citep{2021ApJ...922..269R}. For the highly off-axis BNSs, such as the one under discussion, it is likely that the FRB signal would also have to pass through this component of the ejecta \citep{2020Sci...370.1450D}, which goes against the FRB-GW association. 

%that would enhance the effective electron density estimate and consequently, makes the ejecta mass constraint even more severe.

%Note that we ignore the effects of a toroidal accretion disc which is expected to be formed from the matter ejected due to tidal torques in the merging process. However, this ejecta would also affect the FRB signal as the merger would have large inclination angle if the association is true \citep{2017PhRvD..95b4029D}.

%Finally, several simulations \citep{2013PhRvD..87b4001H} show a decrease in dynamical ejecta mass with a softer equation of state (EOS). 

Simulations such as \citet{2017PhRvD..95b4029D,Camilletti_2022_NRsims} show a decrease in dynamical ejecta mass with a softer equation of state (EOS), at least for asymmetric mass binaries. Therefore a softer EOS would be required in the case, if we consider our updated parameter estimation gives on the mass ratio $q=m_2/m_1=0.63^{+0.32}_{-0.24}$ with the high spin prior. Note that given the large uncertainties, the binary may still be near-equal mass, in which case a stiffer EOS may produce less ejecta than a softer one. On the other hand, in the case of GW190425, the presence of a supramassive neutron star requires a very stiff EOS \citep{2013PhRvL.111m1101B}. 
Additionally, the accuracy and reliability of ejecta mass estimates in grid-based simulations are dependent on the level of spatial resolution achieved in resolving the collision dynamics \citep{2014MNRAS.437L...6K,2017PhRvD..96h4060K}. However, it is not clear if this would decrease the amount of expected ejecta mass in our scenario \citep{2017PhRvD..96h4060K}. For instance, it is worth considering that the presence of a highly magnetized neutron star remnant could potentially lead to an increase in ejecta mass. If the remnant resulting from the binary neutron star merger is a supramassive neutron star, the substantial neutrino luminosity emitted by the remnant itself may contribute to the ejection of a significant mass ($\rm \sim 10^{-3}-10^{-2} M_{\odot}$; \citealt{2009ApJ...690.1681D,2014MNRAS.443.3134P,2015ApJ...813....2M}).
%For example, there are reasons to believe that the presence of a highly magnetized neutron star remnant would in fact increase the ejecta mass. If the remnant of the BNS merger is indeed a supramassive neutron star (SMNS), then it is possible that the large neutrino luminosity from the remnant itself would ejects a non-negligible amount of mass (∼10−3 solar masses; Dessart et al. 2009; Perego et al. 2014; Martin et al. 2015; Richers et al. 2015). 
Additionally, it is expected that the supramassive neutron star would have a strong ordered magnetic field, as seen in cosmological simulations \citep{2022PhRvD.106b3013P}, which can further enhance the outflow \citep{2004ApJ...611..380T,2018ApJ...857...95M}.
%and a requirement in the curvature radiation models (cite paper)
Therefore, we do not expect that the stiff equation of state (EOS) and the limitations in the resolution of numerical simulations would alter our conclusion.

It is worth noting that no kilonova (KN) was detected in association with this GW event (e.g. \citealt{Coughlin_2019,Hosseinzadeh_2019}), although only a fraction ($\sim 40\%$, mostly in the north Galactic Cap) was covered by follow up searches, so a KN may have easily been missed. It is relevant that a KN was not detected at the location of the FRB, which the Zwicky Transient Facility (ZTF) and the Palomar Gattini-IR telescopes imaged \citep{Coughlin_2019}. Those searches were sensitive down to $g$ and $r$ $\sim 21$ mag, and to 15.5 on $J$-band, and a GW170817--like KN, which had ejecta mass similar to that claimed in \citet{moroianu2022assessment}, should have been detected if it was present. While the non-detection may imply lower ejecta mass, as expected for higher mass BNSs, the upper limits are not stringent enough to ensure the extremely low ejecta mass required for the FRB to be detected, hence the KN non-detection is inconclusive for this FRB. 

High-mass mergers such as GW190425 are expected to promptly produce a black hole, resulting in a small amount of ejecta, that are especially rich in lanthanides. This scenario likely results in a faint, particularly red EM counterpart \citep{Foley_2020}, hence explaining the lack of kilonova detection. To check this possibility, we repeat the analysis of \citep{abbott2020gw190425} and compute the probability of prompt collapse as well as the threshold mass $M_{\rm{threshold}}$ for which BNS systems are expected to collapse into a black hole promptly after merger. To estimate the threshold mass, as in \citep{abbott2020gw190425} we consider two cases: 1) using EOS constraints from GW170817 and 2) additionally imposing a maximum TOV mass\footnote{Tolman–Oppenheimer–Volkoff limit} $ M_{\rm{TOV}}^{\rm{max}} \geq 1.97 M_{\odot}$.

\begin{figure}[ht!]
    \centering
    \includegraphics[width=0.49\textwidth]{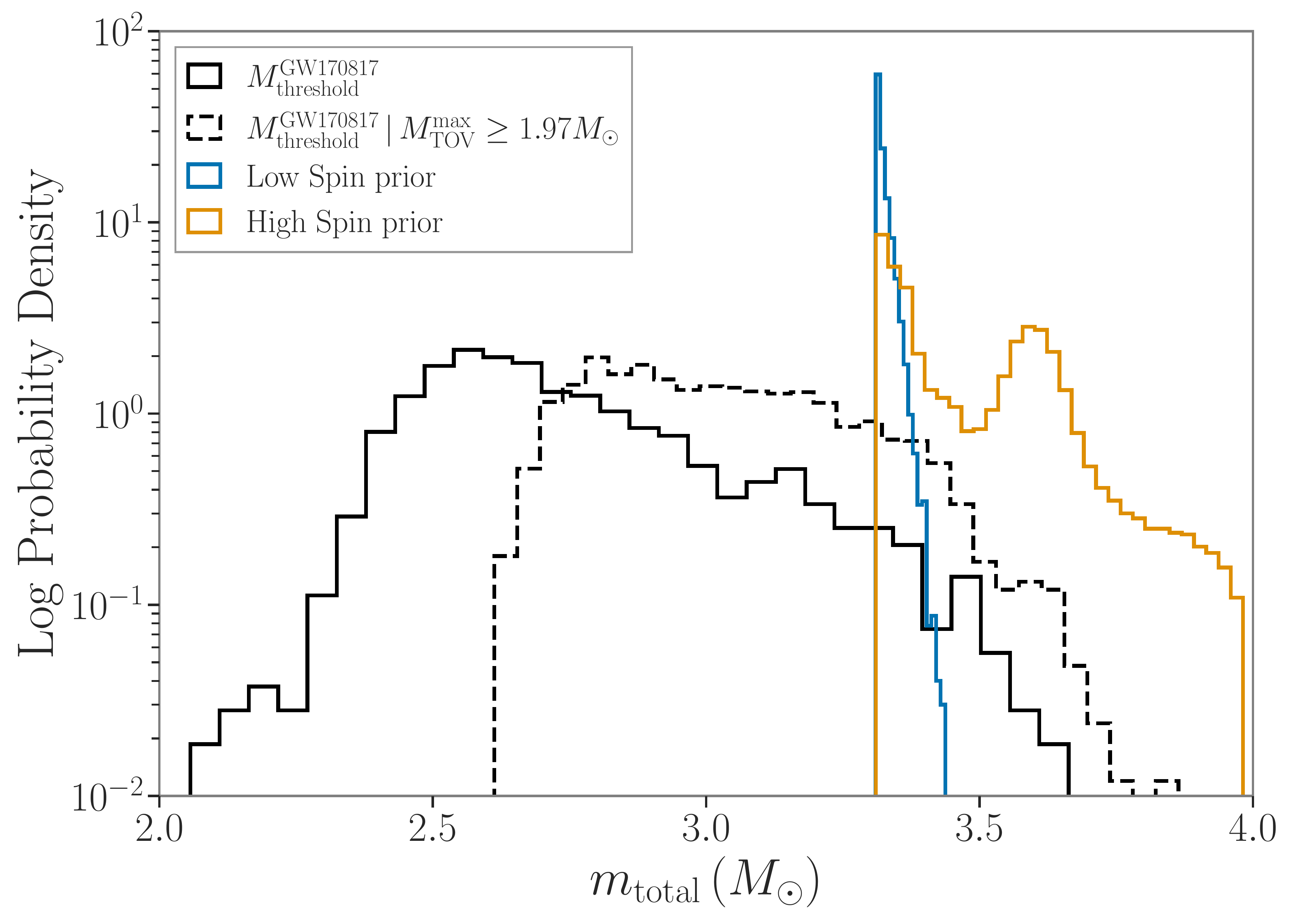}
    \caption{Posterior distributions on the total mass $m_{\rm{tot}}$ for \GW from the fixed 3D sky position of \GAL parameter estimation results for both low (blue) and high spin (yellow) priors. We also show the estimated prompt collapse threshold mass $M_{\rm{threshold}}$ posteriors, conditional on EOS constraints from GW170817 (solid black) and with the additional constraint assuming $M_{\rm{TOV}}^{\rm{max}} \geq 1.97 M_{\odot}$ (dashed black).
    }\label{figure:mth}
\end{figure}

We use the estimated maximum mass for \GW with the fixed position posterior samples to estimate the mass of the resulting NS after merger. First, we note that this updated total mass estimate is still $>3.2~M_\odot$ for all priors considered in this study, therefore still supporting a high mass merger, as in \citet{abbott2020gw190425}. In Figure~\ref{figure:mth}, we show the posterior distribution for the total mass of \GW compared to the inferred $M_{\rm{threshold}}$. 
Following ~\cite{agathos2020inferring}, the prompt collapse probability can be calculated as:
 \begin{align}
     P({\rm PC}_{\rm{GW190425}}|d)&=P(m_{\rm{tot}}>M_{\rm{threshold}}) \nonumber\\
     &=P(m_{\rm{tot}}-M_{\rm{threshold}}>0)
     \end{align}
where $\rm{d}$ is the gravitational-wave observed data.
We find that since the total mass for \GW increases when fixing the position of its host to \GAL, the corresponding probabilities for prompt collapse are above 88\% for all spin priors. Specifically for high spin priors, we find 98\% when assuming EOS constraints from GW170817 only and 94\% by additionally imposing a maximum TOV mass constraint.

\subsection{Implication for future FRB-GW associations} \label{section:frb-model}

This study clearly highlights that any potential association between the FRB and the remnant of a BNS merger, as discussed in \cite{moroianu2022assessment}, must satisfy a range of rigorous constraints beyond the consideration of low chance association probability. One notable constraint is related to the detectability of the FRB at $\sim$ 1 GHz. As detailed in \S\ref{sec:propagation}, the delay after which the FRB can propagate without suffering significant attenuation depends on the ejecta mass. In the case of an ejecta mass as low as $10^{-4}$ M$_{\odot}$, it could take several months ($\sim$ a few million seconds) for the ejecta to become optically thin. However, the stability of SMNS remnants over such a prolonged period seems unlikely \citep{2020GReGr..52..108B,2020JHEAp..27...33L}. Hence,  the fraction of BNS mergers from which we can detect FRBs through the ``blitzar" mechanism is likely very small \citep[$\lesssim$ 1\%;][]{2021ApJ...920..109B}, suggesting a volumetric formation rate of stable neutron star remnants capable of producing FRBs to be $\lesssim$ 10 Gpc$^{-3}$ yr$^{-1}$. If we assume the typical magnetic active timescale of the remnant stable neutron star to be 20 years \citep{2016ApJ...833..261B,2019ApJ...886..110M} and comoving number density of local Universe repeating FRBs $\gtrsim$ 10$^{5}$ Gpc$^{-3}$ \citep{2021ApJ...919L..24B}, it suggests that $<~$1\%  of repeating FRBs can be produced by binary neutron star merger remnants, as also proposed by \cite{2020ApJ...893...44Z}.

One potential solution to circumvent the challenges posed by propagation effects is that the FRB traverses in the direction of the relativistic jet, as discussed in \S\ref{section:related}. In this scenario, one needs to consider only those BNS mergers that also produce short GRBs.
However, we note that it is still unclear if the cavity created by the relativistic jet would remain low-density for an extended period, ranging from hours to days \citep{2020ApJ...900L..35C,2021MNRAS.500..627H}. %Therefore, caution must be exercised in associating FRBs, even with on-axis BNS mergers emitted within a few seconds after short GRBs.
\cite{2022PhRvD.105h3004S} estimated that $\approx$ 2\% of BNS mergers would likely produce an observable sGRB.
Using this, we estimate the fraction of BNS mergers that can be related to non-repeating FRBs. The estimated local volumetric rate of BNS mergers 250$-$2810 Gpc$^{-3}$ yr$^{-1}$ \citep{abbott2020gw190425} is $\approx$ 100 times less than that of FRBs with energy $\geq$ 10$^{39}$ ergs \citep{2023ApJ...944..105S}. Using the sGRB constraint,%that only on$-$axis BNS remnants (viewing angle $\lesssim 10\deg$) that  observable short gamma-ray burst to be 0.02can potentially let the FRB signal escape the ejecta
the difference between FRB and BNS merger volumetric rate would increase by $\sim$ 100 times. 
This difference will further increase if we incorporate the fact that only $\sim$ 45$-$90\% BNS remnants are expected to form supermassive or stable NSs \citep{2021ApJ...920..109B}.
%, and (2) not all binary neutron star mergers can form successful relativistic jets \citep[$\sim 20$\% can;][]{2022A&A...666A.174S}.
%and (3) the fraction of supramassive neutron star that remains stable for only a few 10s of seconds after relativistic jet $<<$ 1. 
 Therefore, the association between FRBs and the post-merger remnants of BNS mergers should be an extremely rare occurrence, if it happens at all, with an exceedingly small fraction of FRBs ($\sim$ 1 in $\sim 10^{4}-10^{5}$ non-repeating FRBs) satisfying the necessary conditions. 

In summary, we argue that BNS merger remnants cannot account for the formation of $>$ 1\% of FRB sources. This implies that short GRBs should not be considered when assessing the overall characteristics of the FRB host population.
% In short, BNS mergers can't be responsible for the formation of $>$ 1\% of the FRB sources. %.and hence, should be used to explain different observed FRB host observables (FRB host offsets, SFRs, stellar mass, etc.)
 %{\color{red} Therefore, BNS mergers and their remnants can only power a very small sub-set of FRB population, rendering claims of similarity between the offsets of FRBs and short GRBs possibly an observational coincidence.}
 %In future work (Bhardwaj et. al, in preparation), we will examine and discuss similar constraints regarding the association of FRBs with BNS mergers for the pre-merger scenario.
%This also suggests that 
%Moreover, any future association should be made using$\sim 5\sigma$ confidence limits after take into account all necessary pre-requisites, such as on-axis, nature of post-merger remnant, and the delay between the two events. This requires that both FRBs and GRBs should at least be confirmed to be located in the same galaxy prior to claiming any association in all future GW-FRB associations.

%Finally, the estimated small volumetric rate of BNS and birth rate of stable neutron stars ($\sim$ 1$-$10 Gpc$^{-3}$ yr$^{-1}$, as discussed above) and expected 

\section{Conclusion} \label{section:conclusion}
In this study, we establish essential criteria rooted in astrophysical and gravitational wave constraints, which must be met by any credible association between FRBs and BNS mergers. These conditions serve as prerequisites for evaluating the likelihood of coincidental occurrences of these two events. To demonstrate the utility of these criteria, we employ the proposed association between the gravitational-wave event GW and FRB, as put forth by \cite{moroianu2022assessment} and \cite{panther2022most}, as a test case.
%In this work, we investigate the association between the gravitational-wave event \GW and \FRB as proposed by \cite{moroianu2022assessment} and \cite{panther2022most}. 
%We have re-calculated the probability of association as a Bayesian hypothesis comparing the hypothesis between a common source for the transients and the chance of a random source association. The posterior odds were calculated, following previous work, as a product of temporal and spatial overlap integrals. We found that the spatial overlap can marginally support a common source hypothesis, yielding a value of $\mathcal{O}(50)$. However, since both the CHIME observatory and the LIGO interferometers point in similar directions, the significance of the spatial overlap is lowered to $\mathcal{O}(10)$. The temporal overlap integrals yield less favourable results since they take into account the correlations between the instruments. The overall posterior odds were found to be $\mathcal{O}(5)$ and to only minimally support the association claimed by \cite{moroianu2022assessment}. 
 We first use the parameter estimation results with the sky location of \GAL as the host galaxy for the \FRB counterpart identified by \cite{panther2022} and \cite{2023arXiv231010018B}, as well as with its measured redshift \citep{gw190425:association}. %This work presents the first comprehensive GW parameter estimation analysis for the claimed associated transients.
 The parameter estimation results yield a stringent constraint on the inclination angle, which strongly contradicts the association hypothesis. In order for the association to be valid within the blitzar model, the GW event needs to be on-axis. However, our findings indicate that the probability  of the viewing angle $p(\theta_{v} > 30^{\circ}) \approx 99.99$\% for both high-spin and low-spin prior scenarios.
%The viewing angle obtained with our parameter estimation results strongly excludes the association hypothesis as for the association to be valid, the authors assumed the GW event to be on-axis, but we find $p(\theta_{v} > 30\deg) >$ 99.98\% in both high-spin and low-spin prior scenarios. 
Furthermore, our analysis demonstrates that in order for the \FRB to be observed at 400 MHz without significant attenuation, an exceedingly low ejecta mass ($\lesssim 10^{-15}$ M$_{\odot}$) is required. This value is orders of magnitude smaller than what is typically anticipated based on simulations and observations of binary neutron star mergers.
%While the measured viewing angle is  consistent with the lack of GRB and kilonova counterparts, the measured total mass of the remnant slightly increases but it's consistent with the results of \cite{abbott2019properties}. 
%\red{As a consequence, the estimated probabilities for prompt collapse into a black hole have increased compared to those first calculated in \cite{abbott2019properties} for both EOS-informed and agnostic priors.}
Additionally, we note that in order for the association to make sense, one would require an exotic equation of state, as suggested by \cite{zhang2022physics}. We, therefore, argue that \GW most likely promptly collapsed into a black hole, a possibility that was also considered by \cite{abbott2019properties}.  Therefore, we conclude that \GW and \FRB are not related. 

Finally, our analysis constrained the volumetric rate of BNS mergers and birth rate of stable neutron star remnants $\lesssim 1-10$ Gpc$^{-3}$ yr$^{-1}$ that can produce FRBs, which is insufficient to explain the high volumetric rate of FRBs. Hence, we conclude that BNS merger remnants cannot account for the formation of $>$ 1\% of FRB sources. Consequently, they should not be invoked to characterize the overall population of FRB host galaxies.

In conclusion, we emphasize the need for caution in associating gravitational wave and fast radio burst events in future studies. Relying on the probability of chance associations can be insufficient to establish potential associations conclusively. This cautionary approach is particularly crucial due to the significant disparity in the local volumetric rates between binary neutron star events, which are approximately four orders of magnitude lower, and FRBs at their fiducial energy of approximately $10^{39}$ ergs. Therefore, it is important to take into account the astrophysical constraints highlighted in this study when considering such associations.
%for the case considered, more observations of potentially associated GW and FRB counterparts will be needed to potentially shed light on the possibility of such transients having a common origin.
%Finally, we compute the Hubble constant assuming the association to be real and show that the value $H_0$ is consistent with what was found using the first bright siren GW170817. To conclude, we bring forward a word of caution when performing electromagnetic counterparts claims.

\section*{Acknowledgements} 
The authors would like to thank Bing Zhang, David Radice, Kendall Ackley, and Pawan Kumar for their useful comments and feedback. MB is a McWilliams postdoctoral fellow and an International Astronomical Union Gruber fellow. IMH is supported by NSF Award No. PHY-1912649 and PHY-2207728. VDE is supported by Science and Technology Facilities Council(STFC) grant ST/V001396/1. The authors are grateful for computational resources provided by the Leonard E Parker Center for Gravitation, Cosmology and Astrophysics at the University of Wisconsin-Milwaukee and supported by NSF awards PHY-1912649, as well as computational resources provided by Cardiff University and supported by STFC grant ST/V001337/1 (UK LIGO Operations award). We thank LIGO and Virgo Collaboration for providing the data for this work. This research has made use of data, software and/or web tools obtained from the Gravitational Wave Open Science Center (https://www.gw-openscience.org/), a service of LIGO Laboratory, the LIGO Scientific Collaboration and the Virgo Collaboration. LIGO Laboratory and Advanced LIGO are funded by the United States National Science Foundation (NSF) as well as the Science and Technology Facilities Council (STFC) of the United Kingdom, the Max-Planck-Society (MPS), and the State of Niedersachsen/Germany for support of the construction of Advanced LIGO and construction and operation of the GEO600 detector. Additional support for Advanced LIGO was provided by the Australian Research Council. Virgo is funded, through the European Gravitational Observatory (EGO), by the French Centre National de Recherche Scientifique (CNRS), the Italian Istituto Nazionale di Fisica Nucleare (INFN) and the Dutch Nikhef, with contributions by institutions from Belgium, Germany, Greece, Hungary, Ireland, Japan, Monaco, Poland, Portugal, Spain. This material is based upon work supported by NSF's LIGO Laboratory which is a major facility fully funded by the National Science Foundation. This article has been assigned LIGO document number LIGO-P2100449.

\bibliography{sample63}{}
\bibliographystyle{aasjournal}

%\appendix

%\section{Can repeating FRBs}
%\label{app:mcmc}

\end{document}